\documentclass[]{aa}

\usepackage{graphicx,natbib}

\def\jh{\mbox{$\rm (J-H)$}}

\def\jk{\mbox{$\rm (J-K_s)$}}

\def\mv{\mbox{$M_{\rm V}$}}
\def\ebv{\mbox{$\rm E(B-V)$}}

\def\aV{\mbox{$A_{\rm V}$}}
\def\rc{\mbox{$\rm R_c$}}
\def\rx{\mbox{$\rm R_{ext}$}}
\def\rhL{\mbox{$\rm R_{hL}$}}
\def\rhSC{\mbox{$\rm R_{hSC}$}}

\def\rh{\mbox{$\rm R_{hl}$}}

\def\rt{\mbox{$\rm R_t$}}

\def\ds{\mbox{$\rm d_\odot$}}
\def\Rgc{\mbox{$\rm R_{GC}$}}
\def\feh{\mbox{$\rm [Fe/H]$}}
\def\jj{\mbox{$\rm J$}}
\def\hh{\mbox{$\rm H$}}
\def\ks{\mbox{$\rm K_s$}}
\def\tdis{\mbox{$t_{\rm dis}$}}

\begin{document}

\title{Structural parameters of 11 faint Galactic globular clusters derived with 2MASS}

\author{C. Bonatto\inst{1} \and E. Bica\inst{1}}

\institute{Universidade Federal do Rio Grande do Sul, Departamento de Astronomia\\
CP\,15051, RS, Porto Alegre 91501-970, Brazil\\
\email{charles@if.ufrgs.br, bica@if.ufrgs.br}
\mail{charles@if.ufrgs.br} }

\date{Received --; accepted --}

\abstract
{Structural parameters and the total \mv\ magnitude are important properties to be used in the
characterization of individual globular clusters. With the increasing statistics, especially
of the intrinsically faint objects, collective properties of the Galactic globular cluster 
system will be better defined, with reflexes on our understanding of the formation history 
of the Galaxy.}
{The determination of structural parameters of 11 faint Galactic globular clusters that,
in most cases, had not been previously studied in this context. The clusters are
IC\,1257, Lyng\aa\,7, Terzan\,4, Terzan\,10, BH\,176, ESO\,452-SC11, ESO\,280-SC08,
2MASS-GC01, 2MASS-GC02, GLIMPSE-C01 and AL\,3, which are projected not far from the central
region of the Galaxy. Field-star contamination is significant in the colour-magnitude diagrams.
Half of the sample have an absorption $\aV\ga7$, reaching in some cases $\aV\ga15$.}
{Stellar radial number-density and surface-brightness profiles are built with 2MASS photometry
that, for the present clusters, corresponds basically to giant-branch stars. Field-star 
decontamination is essential for clusters in dense fields, so an algorithm that we previously 
developed for open clusters in rich fields is employed to better define cluster sequences. 
With decontaminated photometry we also compute the total \mv\ of four such globular clusters, 
using M\,4 as a template. King-like functions are fitted to the radial profiles, from which the 
core, half-light, half-star count and tidal radii are derived, together with the concentration 
parameter. Parameters derived here are compared to the equivalent ones of other Galactic
globular clusters available in the literature.}
{Compared to massive globular clusters, those in the present sample have smaller tidal and larger 
core radii for a given Galactocentric distance, which is consistent with rather loose structures. 
Globular cluster radii derived from number-density and surface-brightness profiles have similar 
values. The present magnitude estimates are $\mv\approx-4.9$ (ESO\,280-SC08), $\mv\approx-5.8$ 
(2MASS-GC01) and $\mv\approx-5.6$ (2MASS-GC02). We also estimate $\mv\approx-6.1$ for GLIMPSE-C01, 
which results somewhat less luminous than previously given. The density profiles of Tz\,4 and 
2MASS-GC01 present evidence of post-core collapse clusters.}
{Structural parameters and luminosity of most of the faint globular clusters dealt with in this
paper are consistent with those of Palomar-like (low-mass and loose structure) globular clusters.
This work helps to improve coverage of the globular cluster parameter space.}


\titlerunning{2MASS structural parameters of faint Galactic GCs}

\authorrunning{C. Bonatto \and E. Bica}

\maketitle

\section{Introduction}
\label{intro}

Astrophysical parameters of globular clusters (GCs) are of great interest, first because they can be
used to characterize GCs individually, what eventually will lead to a deeper understanding of their
properties as a class in the Galaxy. On a broad perspective, such parameters may provide important
clues to the investigation of the formation and evolution processes of the GC system of the Galaxy
and elsewhere, e.g. Harris (1996, and the 2003 update\footnote{\em http://physun.physics.mcmaster.ca/Globular.html}
- hereafter H03), \citet{MvdB05} and \citet{GCProp}. On the observational perspective, the depth of
this kind of analysis depends directly on the statistical completeness of the sample, especially in
the faint-GC tail, and the reliability of the derived parameters.

In general terms, the standard picture of the GC structure assumes a isothermal central region
and a tidally truncated outer region, usually described by a roughly spherical symmetry (e.g. 
\citealt{Binney1998}). In this context, the structure of most GCs is well approximated by a rather 
dense core and a sparse halo, but with a broad range in concentration. The most commonly used 
structural parameters are the core (\rc), half-light (\rhL) and tidal radii (\rt), as well as the
concentration parameter $c=\log(\rt/\rc)$, related to the isothermal sphere, single-mass model 
introduced by \citet{King1962} to fit the surface brightness profiles (SBPs) of Galactic GCs.

As they age, GCs are continually harassed by external processes such as tidal stress and dynamical
friction (from the Galactic bulge and disk, and giant molecular clouds), and internal ones such as
mass loss by stellar evolution, mass segregation and low-mass star evaporation (e.g. \citealt{Khalisi07};
\citealt{Lamers05}; \citealt{GnOs97}). Over a Hubble time, these effects tend to decrease cluster mass,
which may accelerate the core collapse phase for some clusters (\citealt{DjMey94}, and references therein).
As a dramatic end-result, low-mass clusters may disrupt completely over the course of a few disruption
time scales (\tdis). Observational evidence for the dependence of \tdis\ on cluster mass are given by,
e.g. \citet{JA82}, \citet{Wielen91} and \citet{Lamers05}. The latter authors found $\tdis\sim M^{0.62}$,
which agrees with the N-body estimate of \citet{BM03} for the disruption time scale due to the Galactic
tidal field. Theoretically, the dependence of \tdis\ on cluster mass was investigated by, e.g.
\citet{Spitzer58} and \citet{GnOs97}. Most of the above processes are related to the large-scale mass
distribution of the Galaxy. Thus, their strength should depend on Galactocentric distance, and they are
expected to be more effective on the less-massive clusters (e.g. \citealt{DjMey94}; \citealt{vdBergh91};
\citealt{CheDj89}). Consequently, the distribution of structural parameters among the Galactic GC population,
as well as their dependence on Galactocentric distance, may also be related to the physical conditions
prevailing at the early formation phase (e.g. \citealt{vdBergh91}).

All these arguments considered, it is natural to assume that the long-term (more than 10\,Gyr, in most
cases) dynamical evolution of the collective GC population (as well as the individual GCs) may be reflected
on the statistical properties of their structural parameters. It is in this context that intrinsically faint
GCs play an important r\^ole, since they are more sensitive to the processes discussed above, and have shorter
dynamical-evolution time scales than the massive ones.

In most cases, SBPs of faint GCs may contain significant noise owing to the intrinsically-low surface
brightness outside the central region, and  because of the random presence of relatively bright stars,
either from the field or cluster members. The same applies to the outer parts, where background starlight
is usually a major contributor even in luminous GCs.  Structural parameters measured in such SBPs would
certainly be affected in varying degrees. In such cases, the alternative is to work with stellar radial
density profiles (RDPs), in which only the projected number-density of stars is taken into account,
irrespective of the individual stellar brightness. RDPs are particularly appropriate for the outer parts,
provided that a statistically significant, and reasonably uniform, comparison field is available to tackle
the background contamination. Such wide fields are provided by 2MASS (e.g. \citealt{DetAnalOCs}; \citealt{BB07a}).
However, the 2MASS photometric depth limit implies that basically the giant branch can be accessed in GCs more
distant than a few kpc from the Sun (\citealt{BB07b}). Thus, it is necessary to take into account depth-limited
effects when considering such structural parameters in the context of absolute comparisons with parameters
obtained with deeper photometry. This issue has been studied in \citet{BB07b} by means of model GCs built
assuming different conditions, which allowed us to investigate relations among radii measured in RDPs and
SBPs, together with their dependence on photometric depth.

\citet{TKD95} presented a catalogue of SBPs for 125 Galactic GCs with structural parameters derived
mostly with \citet{King66} profile. Based mostly on the latter work, H03 provided structural parameters
for 141 (of the total catalogued 150 at the time) GCs. Recently, \citet{Cohen07} used surface brightness
profiles in \jj, \hh\ and \ks\ with 2MASS\footnote{{\em http://www.ipac.caltech.edu/2mass/releases/allsky/}}
photometry to derive core radii for 104 GCs, by means of fits of a power-law with a core function.
The 9 remaining GCs in H03 are faint and were not included in \citet{Cohen07}.

The compilation of H03 contains 150 GCs but, as a consequence of recent deeper surveys, the number of known
GCs in the Galaxy has been increasing. For instance, \citet{KMB2005} discovered the far-IR GC GLIMPSE-C01.
\citet{Carraro05} identified Whiting\,1 as a young halo GC. Two stellar systems detected with the Sloan
Digital Sky Survey (SDSS) in the outer halo, Willman\,1 (SDSS\,J1049$+$5103) and SDSS\,J1257$+$3419, might
be GCs or dwarf galaxies (\citealt{Willman05}; \citealt{Sakamoto06}). \citet{OBB06} identified AL\,3 as a
new GC in the bulge with a blue horizontal branch. Next, \citet{FMS07} found evidence that FSR\,1735 is an
inner Galaxy GC, and \citet{Belokurov07} found the faint halo GC SEGUE\,1 using SDSS. More recently,
\citet{Koposov07} reported the discovery of two very-low luminosity halo GCs (Koposov\,1 and 2) detected with
SDSS. The latter three GCs are beyond the scope of the present paper because they are too distant and/or
exceedingly faint. Another object, FSR\,0584, was recently identified as a nearby Palomar-like halo GC or an
old open cluster (OC) by \citet{FSR584}. Finally, \citet{FSR1767} identified FSR\,1767 as a bulge-projected
($\ell=352.6^\circ$, $b=-2.17^\circ$) Palomar-like GC ($\mv\approx-4.7$) that appears to be the nearest one
($\ds\approx1.5$\,kpc) so far discovered. Fundamental parameters and updates to the GCs in H03 were provided
by \citet{GCProp}, also available as a Vizier on-line
catalogue\footnote{\em http://vizier.cfa.harvard.edu/viz-bin/VizieR?-source=J/A$+$A/450/105}.

The main goal of the present work is to derive structural parameters, using RDPs and SBPs built with 2MASS
star-counts and photometry, for the remaining 9 faint GCs in H03, which lack this kind of data. The targets
are IC\,1257, Lyng\aa\,7, Terzan\,4, Terzan\,10, BH\,176, ESO\,452-SC11, ESO\,280-SC08, 2MASS-GC01, and
2MASS-GC02. We also include in the analyses the more recently discovered GCs GLIMPSE-C01 (\citealt{KMB2005})
and AL\,3 (\citealt{OBB06}).

This paper is structured as follows. In Sect.~\ref{TargetGCs} we present previous results on the 
present GCs that are relevant for this work. In Sect.~\ref{PhotPar} we detail the 2MASS photometry,
including extraction and errors, build the near-infrared colour-magnitude diagrams (CMDs) for the 
present GCs, briefly discuss the tools and algorithms employed to decontaminate the CMDs of field
stars, and estimate the absolute $V$ magnitudes of ESO\,280-SC06, 2MASS-GC01, 2MASS-GC02 and
GLIMPSE-C01. In Sect.~\ref{Struc} we build RDPs and SBPs and derive the core, half-light, half-star 
count and tidal radii, together with the concentration parameter for the present GC sample. In
Sect.~\ref{Disc} we discuss potential effects of depth-limited photometry in the derivation of 
structural parameters from radial profiles, and compare such parameters with those of other globular 
clusters in the Galaxy. Finally, concluding remarks are given in Sect.~\ref{Conclu}.

\section{The faint GC sample}
\label{TargetGCs}

In the following we recall fundamental literature results for the present analyses.

\paragraph{\tt IC\,1257:} This object was first classified as an OC (\citealt{L87}
and references therein), but \citet{H97} identified it as a GC using B, V and I photometry.
They also determined the cluster reddening $\ebv=0.75$, distance from the Sun $\ds=24$\,kpc
and metallicity $\feh=-1.7$. Its absolute magnitude is $\mv=-6.15$ (H03).
 
\paragraph{\tt Lyng\aa\,7:} \citet{L64} discovered and classified Lyng\aa\,7 as a Trumpler type
II\,2\,p OC, which was later reported in \citet{Alter70}. \citet{OBB93} identified it as a GC with
B, V and I photometry, and determined $\ebv=0.72$, $\feh=-0.40$ and $\ds=6.7$\,kpc. \citet{TavaFriel95}
using moderate resolution spectroscopy obtained a radial velocity $\rm V_R=6\pm15\,km\,s^{-1}$, 
$\feh=-0.62$, $\ebv=0.70$ and $\ds=7.2$\,kpc. Based on 2MASS photometry, \citet{Saraj04} derived
$\feh=-0.76$ and $\ds=7.3$\,kpc adopting $\ebv=0.73$.

\paragraph{\tt Terzan\,4:} This GC was discovered by \citet{Terzan71}. Based on V, I and
Gunn Z photometry, \citet{OBiB97} derived $\ebv=2.35$, $\ds=8.3$\,kpc and $\feh=-2.0$.
\citet{OriRich04} obtained $\feh=-1.6$ using high dispersion spectroscopy. \citet{Orto07}
derived $\ds=8.0$\,kpc from HST-NICMOS photometry. Its absolute magnitude is $\mv=-6.09$
(H03).

\paragraph{\tt Terzan\,10:} Another GC discovered by \citet{Terzan71}. \citet{OBiB97}
derived $\ebv=2.40$, $\ds=4.8$\,kpc and $\feh=-1.0$ from V, I and Gunn Z photometry.
Its absolute magnitude is $\mv=-6.31$ (H03).
 
\paragraph{\tt BH\,176:} It was discovered by \citet{BH75}. \citet{OBB95} obtained $\ebv=0.77$,
$\ds=13.4$\,kpc and a nearly solar metallicity by means of V and I photometry. They emphasized
that owing to its low Galactic latitude and high metallicity, the nature of BH\,176 as a metal-rich
GC or an old OC was not clear. \citet{PhSch03} obtained an age of 7\,Gyr, $\ebv=0.53$, $\ds=18$\,kpc,
and $\feh=0.0$ with V and I photometry. They pointed out that that the cluster might be an old
metal-rich OC or a young metal-rich GC. Another possibility is that BH\,176 is associated with an
accreted galaxy event now related with the Galactic anticenter stellar structure or the Monoceros
Ring (\citealt{Frinch06}). H03 provides the absolute magnitude $\mv=-4.35$.

\paragraph{\tt ESO\,452-SC11:} It was discovered and classified as a GC by \citet{Lauberts81}.
\citet{BOB99} confirmed it as a GC and determined $\ebv=0.58$, $\ds=6.5$\,kpc and $\feh=-1.5$
from V and I photometry. Using V and I photometry \citet{Cornish06} found $\ebv=0.57$, $\ds=7.1$\,kpc
and $\feh=-1.15$. ESO\,452-SC11 is intrinsically faint, with $\mv=-3.97$ (H03).

\paragraph{\tt ESO\,280-SC06:} \citet{Lauberts82} reported and classified ESO\,280-SC06 as an
OC. \citet{OBB00} identified it as a GC, and also obtained $\ebv=0.07$, $\feh=-1.8$, and
$\ds=21.9$\,kpc from V and I photometry.

\paragraph{\tt 2MASS-GC01:} The infrared 2MASS-GC01 was discovered by \citet{Hurt00}.
\citet{Ivanov00} obtained $\ebv=6.74$ and $\ds=3.1$\,kpc from  NTT near IR photometry.
\citet{IvaBori02} obtained $\feh=-1.19$ from 2MASS photometry.

\paragraph{\tt 2MASS-GC02:} The infrared GC 2MASS-GC02 was discovered by \citet{Hurt00}.
\citet{Ivanov00} obtained $\ebv=5.55$, $\ds=3.9$\,kpc and $\feh=-0.66$ from  NTT near IR
photometry.

\paragraph{\tt GLIMPSE-C01:} The infrared GC GLIMPSE-C01 was discovered by \citet{KMB2005}
who obtained a kinematic distance of $3.1 - 5.2$\,kpc, $\rc=30\arcsec$ and $\rh=36\arcsec$,
and estimated a total magnitude $\mv=-8.4\pm3$. \citet{IKB05} derived $\ebv=5.0$, $\ds=3.7$\,kpc
and $\feh=-1.61$ from NTT near IR photometry. They also obtained the structural parameters
$\rc=0.84$\,pc ($\approx0.76\arcmin$) and $\rt=29$\,pc ($\approx26\arcmin$) which we compare 
with the present values (Sect.~\ref{Struc}). These cluster radii (especially the tidal) were 
estimated by extrapolation, since they worked with NTT/Sofi photometry, that covers an area of
$4.92\arcmin\times4.92\arcmin$ ($\rm\approx5\times5\,pc^2$), which does not reach to the tidal
radius.

\paragraph{\tt AL\,3:} This cluster was discovered by \citet{AndLind67} and was also catalogued as
BH\,261 by \citet{BH75}. In these studies it was classified as an OC. \citet{OBB06} identified it
as a GC with B, V and I photometry and derived $\ebv=0.36$, $\ds=6.0$\,kpc, $\feh=-1.3$, a half-density
radius $R_{hd}=2.2$\,pc and $\mv=-4.0$.

Table~\ref{tab1} summarizes the fundamental data for the sample GCs. It also contains absolute
magnitude estimates derived in this work (Sect.~\ref{AbsMag}) for some GCs that lacked this
information. For GLIMPSE-C01 we include an additional row that gives our estimate of \mv.

\begin{figure}
\resizebox{\hsize}{!}{\includegraphics{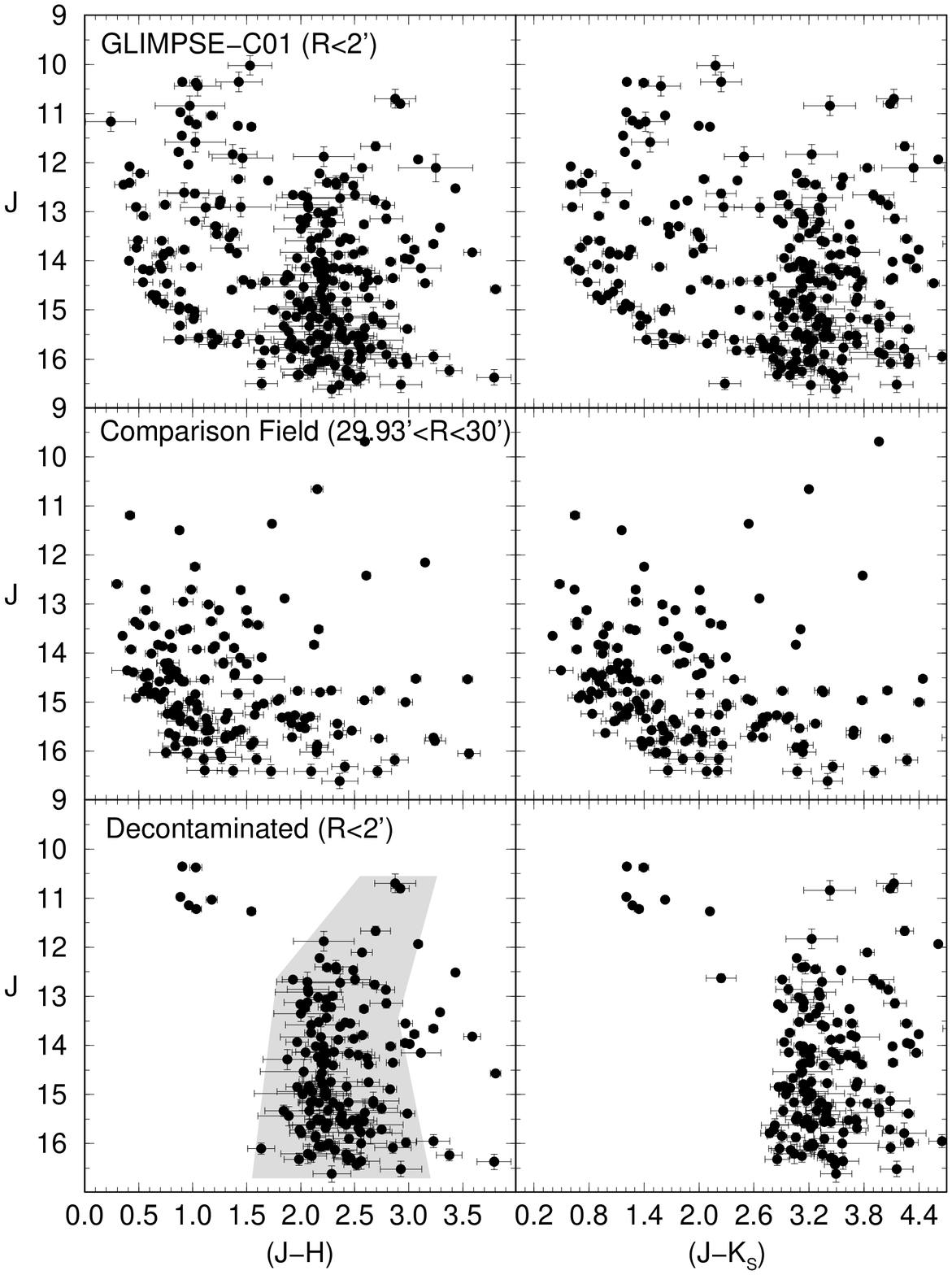}}
\caption{2MASS CMDs extracted from the $R<2\arcmin$ region of GLIMPSE-C01. Top panels: observed 
photometry with the colours $\jj\times\jh$ (left) and $\jj\times\jk$ (right). Middle: equal-area 
comparison field showing that the dominant contamination, in this case, arises from disk stars. 
Bottom panels: field star decontaminated CMDs. The colour-magnitude filter used to isolate probable 
cluster stars (Sect.~\ref{CMF}) is shown as a shaded region.}
\label{fig1}
\end{figure}

\begin{figure}
\resizebox{\hsize}{!}{\includegraphics{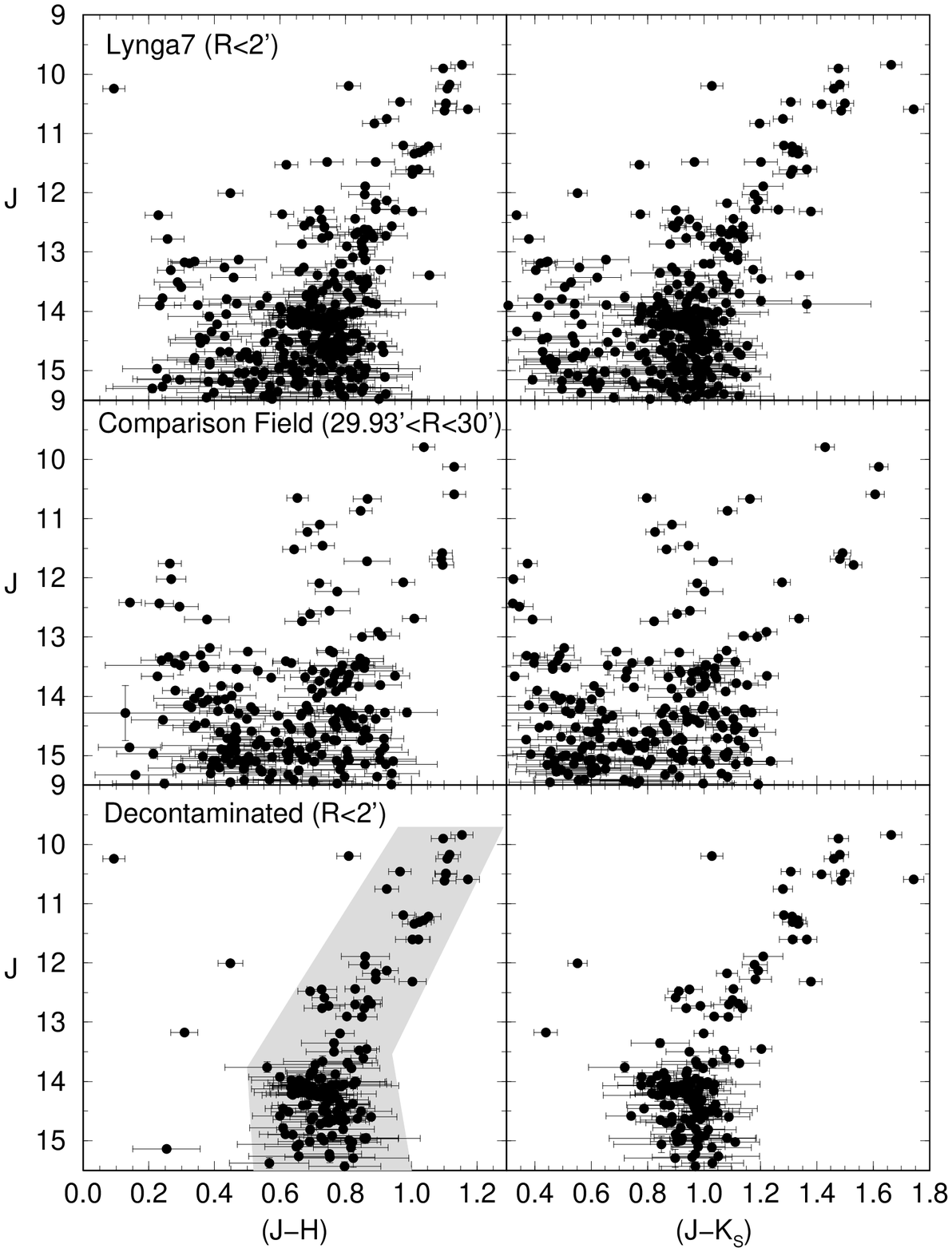}}
\caption{Same as Fig.~\ref{fig1} for the $R<2\arcmin$ region of Lyng\aa\,7.}
\label{fig2}
\end{figure}

\begin{figure}
\resizebox{\hsize}{!}{\includegraphics{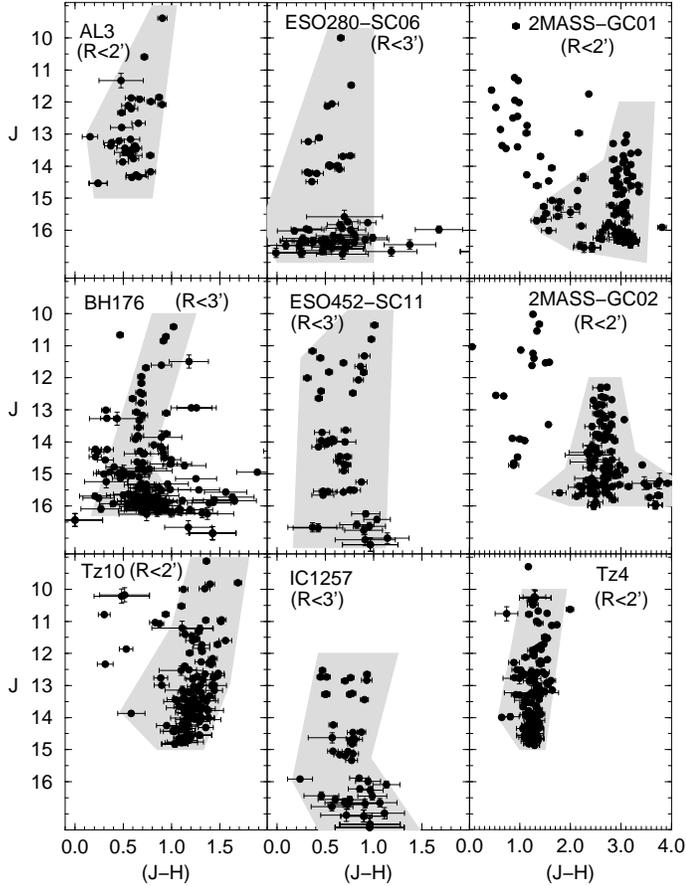}}
\caption{Decontaminated $\jj\times\jh$ CMDs of the remaining GCs. The colour-magnitude filters 
are shown as a shaded region.}
\label{fig3}
\end{figure}

\begin{table*}
\caption[]{Fundamental parameters in the literature and 2MASS extraction radius}
\label{tab1}
\renewcommand{\tabcolsep}{2.6mm}
\renewcommand{\arraystretch}{1.2}
\begin{tabular}{lccrrrcccrr}
\hline\hline
GC&$\alpha(2000)$&$\delta(2000)$&$\ell$&$b$&\rx&\mv&\ebv&\feh&\ds&\Rgc\\
&(hms)&($^\circ\arcmin\arcsec$)&($^\circ$)&($^\circ$)&(\arcmin)& &&&(kpc)&(kpc)\\
(1)&(2)&(3)&(4)&(5)&(6)&(7)&(8)&(9)&(10)&(11)\\
\hline
IC\,1257     &17:27:58&$-$07:05:35& 16.53& $+15.15$& 60&$-6.15^{**}$ & 0.73 & $-1.70$ &25.0 &18.5 \\
Lyng\"a\,7   &16:11:03&$-$55:18:52&328.77& $-2.79$ & 40&$-6.37^\dagger$ & 0.73 & $-0.62$ & 7.2 & 3.9 \\
Tz\,4        &17:30:34&$-$31:35:44&356.02& $+1.31$ & 40&$-6.09^{**}$ & 2.35 & $-1.60$ & 9.1 & 2.0 \\
Tz\,10       &18:02:57&$-$26:04:00& 4.42 & $-1.86$ & 40&$-6.31^{**}$ & 2.40 & $-1.00$ & 5.7 & 1.6 \\
BH\,176      &15:39:07&$-$50:03:02&328.41& $+4.34$ & 40&$-4.35^{**}$ & 0.77 & $~~~0.00$ &15.6 & 10.2\\
ESO\,452-SC11&16:39:26&$-$28:23:52&351.91& $+12.10$& 60&$-3.97^{**}$ &0.49 & $-1.50$ & 7.8 & 2.0 \\
ESO\,280-SC06&18:09:12&$-$46:24:00&346.93& $-12.58$& 50&$-4.9\pm0.3^*$& 0.07 & $-1.80$ &21.7 &15.0 \\
2MASS-GC01   &18:08:22&$-$19:49:47&10.47 & $+0.10$ & 40&$-5.8\pm0.2^*$& 6.80 & $-1.19$ & 3.1 & 4.2\\
2MASS-GC02   &18:09:37&$-$20:46:44& 9.78 & $-0.62$ & 40&$-5.6\pm0.2^*$& 5.56 & $-0.66$ & 3.9 & 3.4\\
GLIMPSE-C01  &18:48:51&$-$01:29:50&31.30 & $-0.10$ & 60&$-8.4\pm3^\ddagger$ & 5.00 & $-1.60$ & 3.7 & 4.5\\
GLIMPSE-C01  &        &           &      &         &   &$-6.1\pm0.2^*$ &      &         &     &     \\
AL\,3        &18:14:06&$-$28:38:12& 3.36 & $-5.27$ & 40&$-4.00^{**}$ & 0.36 & $-1.30$ & 6.0 & 1.4\\
\hline
\end{tabular}
\begin{list}{Table Notes.}
\item Cols.~2 and 3: Central coordinates from H03. Col.~6: 2MASS extraction radius. (**): \mv\ 
from H03; (*): \mv\ estimated by comparison with M\,4 (Sect.~\ref{AbsMag});
$(\dagger)$: \mv\ from \citet{GCProp}. $(\ddagger)$: \mv\ from \citet{KMB2005}. Col.~10:
distance from the Sun from \citet{GCProp}. Col.~11: Galactocentric distance computed using 
$\Rgc=7.2$\,kpc (\citealt{GCProp}) as the distance of the Sun to the Galactic centre. All
additional information from H03.
\end{list}
\end{table*}

\section{2MASS photometry and CMD analytical tools}
\label{PhotPar}

In recent years we developed a series of tools to statistically disentangle field and cluster stars that
have been applied to objects in a wide variety of environmental conditions. They are based on wide-field
2MASS photometry and essentially deal with CMDs and stellar RDPs. As a result, CMDs containing more enhanced
cluster sequences, and RDPs with significant contrast with the background have been obtained, which in turn,
allowed us to derive more constrained cluster fundamental and structural parameters. For instance, nearby OCs
were analysed in detail like NGC\,3960 and M\,52 (\citealt{N3960}), and NGC\,4755 (\citealt{N4755}), embedded
clusters like NGC\,6611 (\citealt{N6611}), and relatively distant OCs like BH\,63 (\citealt{FaintOCs}). The
tools are detailed in \citet{BB07a}, where we studied relatively distant OCs in heavily contaminated fields
towards the bulge and/or low-disk directions.

With respect to cluster structure, most of our analysis is based on large-scale radial profiles, which are
fundamental to derive structural parameters of star clusters, the tidal radius in particular. Such profiles
can be obtained using the uniform, all-sky photometric coverage provided by 2MASS (e.g. \citealt{BB07a};
\citealt{DetAnalOCs}).

\subsection{2MASS extractions}
\label{2massExtr}

The present faint GCs were analysed with \jj, \hh\ and \ks\ 2MASS photometry extracted in circular 
fields of radius \rx\ (Table~\ref{tab1}) centred on the coordinates of the objects (Table~\ref{tab1})
using VizieR\footnote{\em vizier.u-strasbg.fr/viz-bin/VizieR?-source=II/246}. Our experience with
OC analysis (e.g. \citealt{BB07a}, and references therein) shows that as long as no other populous
cluster is present in the field, and differential absorption is not prohibitive, wide extraction 
areas can provide the required stellar statistics, in terms of magnitude and colours, for a consistent 
field star decontamination (Sect.~\ref{FSD}). Working with wide comparison fields also results in stellar
radial profiles with more enhanced contrast with the background (Sect.~\ref{Struc}).

Based on our experience with the analysis of clusters in crowded field directions using 2MASS photometry
(\citealt{BB07a}, and references therein), the upper limit for the \jj, \hh\ and \ks\ photometric 
uncertainties in the 2MASS extractions was adopted as 0.5\,mag. This condition represents a compromise 
between statistically significant star-counts and photometric quality. Note that about $75\%-85\%$ of the
extracted stars have errors smaller than 0.06\,mag in the three 2MASS bands. A typical distribution of 2MASS
errors as a function of magnitude, for clusters at approximately the same central directions as the present
GCs, can be found in \citet{BB07a}. The relatively small size of error bars, propagated to the colours, can
be appreciated in the actual CMDs of the present sample GCs shown in Figs.~\ref{fig1} - \ref{fig3}.

We illustrate as CMD analyses GLIMPSE-C01 (Fig.~\ref{fig1}) and Lyng\aa\,7 (Fig.~\ref{fig2}), which
are projected $\approx30^\circ$ from the direction of the center (Table~\ref{tab1}). 2MASS $\jj\times\jh$
and $\jj\times\jk$ CMDs extracted from a central ($R<2\arcmin$) region of both GCs (top panels) can be
contrasted with the respective equal-area CMDs extracted from the comparison field (middle panels). It is
clear that in both cases the main contaminant are disk stars, with some contribution from bulge stars.

\begin{figure}
\resizebox{\hsize}{!}{\includegraphics{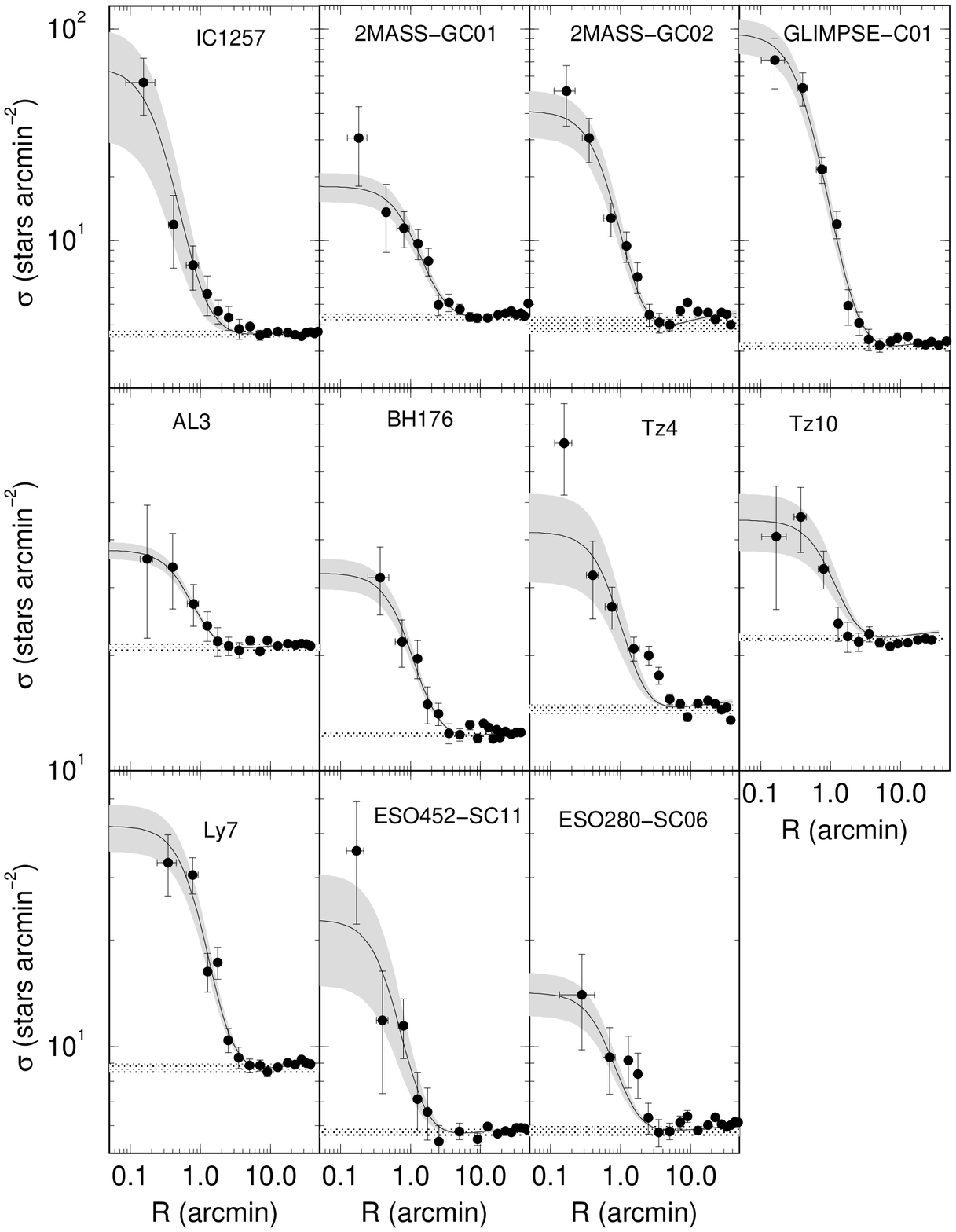}}
\caption{Stellar RDPs (filled circles) in angular scale. Solid lines: best-fit King-like
profile (Eq.~\ref{eq1}). Horizontal shaded region: background level ($\sigma_{bg}$). Gray
regions: $1\sigma$ fit uncertainty range.}
\label{fig4}
\end{figure}

\begin{figure}
\resizebox{\hsize}{!}{\includegraphics{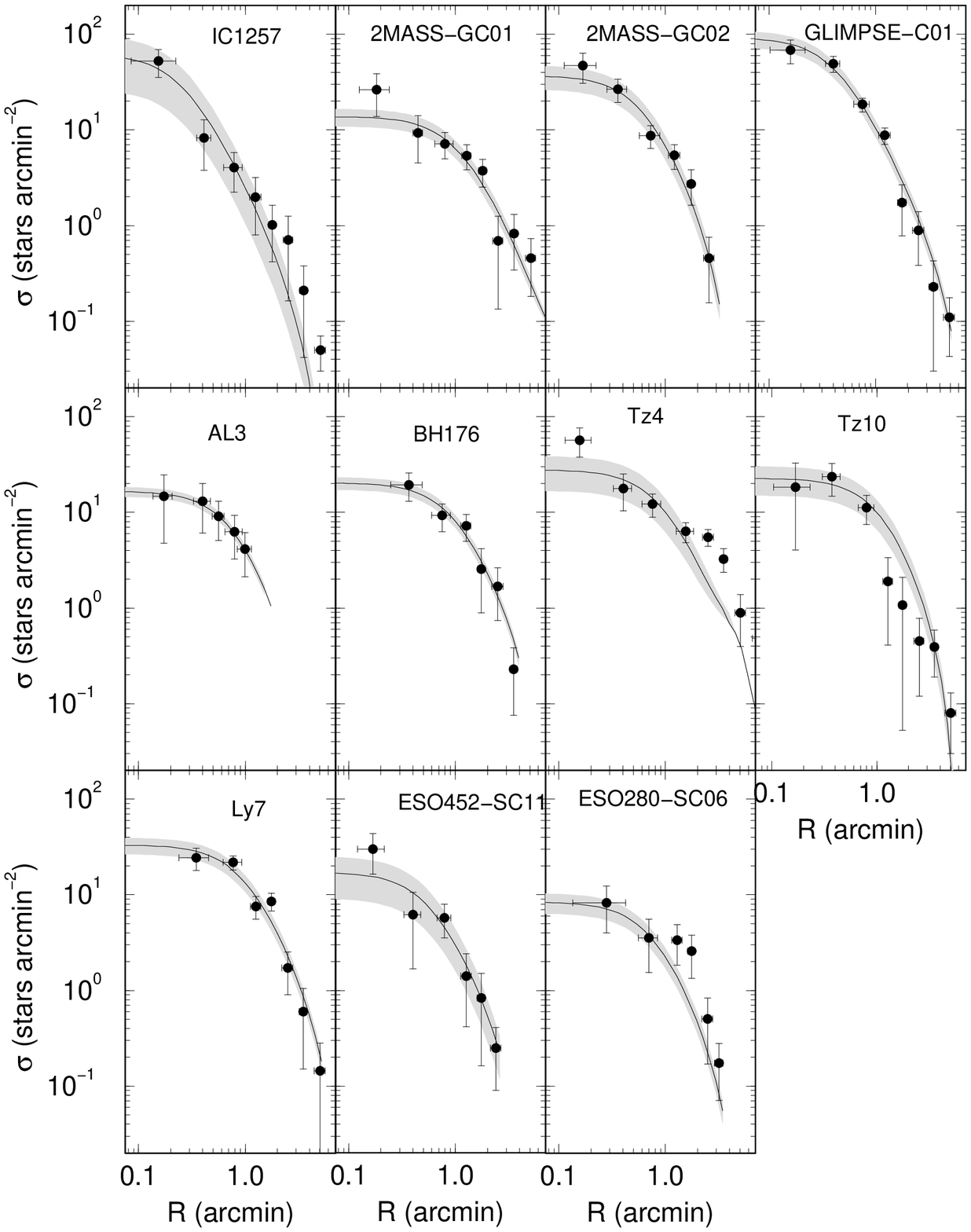}}
\caption{Same as Fig.~\ref{fig4} for the background-subtracted RDPs.}
\label{fig5}
\end{figure}

\subsection{Field-star decontamination}
\label{FSD}

To disentangle the intrinsic CMD morphology from the field we apply the statistical decontamination
algorithm described in \cite{BB07a}. It measures the relative number-densities of candidate cluster and
field stars in small cubic CMD cells with axes corresponding to the magnitude \jj\ and the colours \jh\
and \jk\ (considering also the $1\sigma$ uncertainties in the 2MASS bands). These colours provide
the maximum discrimination among CMD sequences for star clusters of different ages (e.g. \citealt{TheoretIsoc}).

Basically, the algorithm {\em (i)} divides the full range of magnitude and colours of a CMD into a
3D grid, {\em (ii)} computes the expected number-density of field stars in each cell based on the number
of comparison field stars with magnitude and colours compatible with those in the cell, and {\em (iii)}
subtracts the expected number of field stars from each cell. Typical cell dimensions
are $\Delta\jj=0.5$, and $\Delta\jh=\Delta\jk=0.25$, which are large enough to allow sufficient star-count
statistics in individual cells and small enough to preserve the morphology of the CMD evolutionary
sequences. As comparison field we use the region within the respective tidal (Table~\ref{tab2}) and 
extraction (Table~\ref{tab1}) radii to obtain representative background statistics.

The algorithm is applied for 3 different grid specifications in each dimension to minimize potential
artifacts introduced by the choice of parameters, thus resulting in 27 different outputs. The average
number of probable cluster stars $\langle N_{cl}\rangle$ is computed from these outputs. Typical standard
deviations of $\langle N_{cl}\rangle$ are at the $\approx2.5\%$ level. The final field star-decontaminated
CMD contains the $\langle N_{cl}\rangle$ stars with the highest number-frequencies. Stars that remain after
the decontamination are in cells where the stellar density presents a clear excess over the
field. Consequently, they have a significant probability of being cluster members. Further details on the
algorithm, including discussions on subtraction efficiency and limitations, are given in \citet{BB07a}.

The resulting decontaminated CMDs of GLIMPSE-C01 and Lyng\aa\,7 are shown in the bottom panels of
Figs.~\ref{fig1} and \ref{fig2}. As expected, most of the disk component is removed, leaving a
sequence of stars that corresponds essentially to the giant branch.

\begin{figure}
\resizebox{\hsize}{!}{\includegraphics{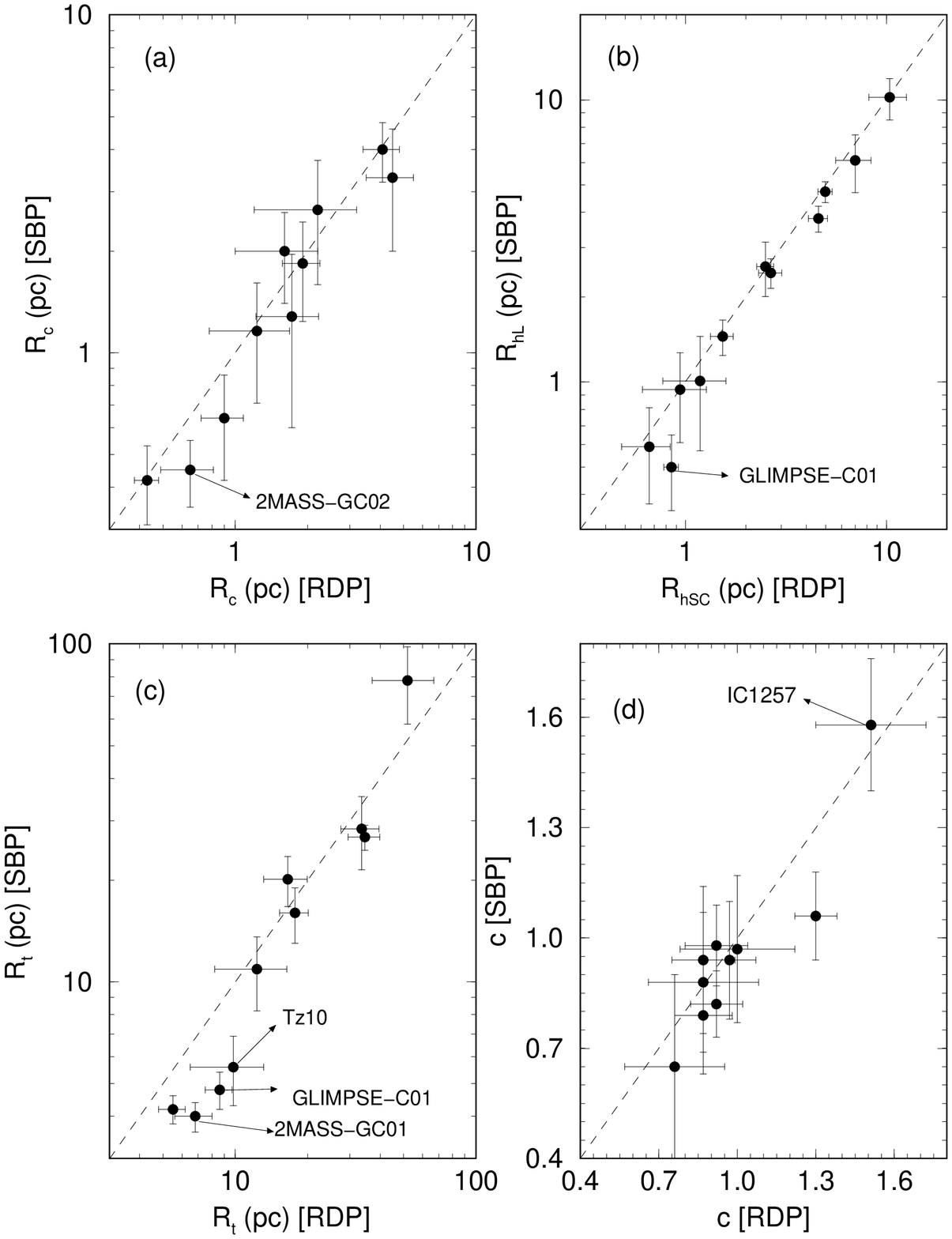}}
\caption{Comparison of cluster structural parameters derived from RDPs and SBPs. Identity is indicated by
the dashed line. GCs that lie at more than $\rm\sim1\sigma$ from the identity line are labelled. The 
highest concentration level (panel d) occurs for IC\,1257, the most distant GC of the present sample, with
$\ds\approx25$\,kpc and $\Rgc\approx18$\,kpc. Absolute units are used.}
\label{fig6}
\end{figure}

For the sake of space, only the decontaminated $\jj\times\jh$ CMDs of the remaining GCs are shown 
in Fig.~\ref{fig3}. In all cases, the spatial region shown in the CMDs lies within the respective 
half-star count radius (Table~\ref{tab2}).

Differential absorption appears to be similar both in each GC and respective field extractions, 
as indicated, for instance, by GLIMPSE-C01 (Fig.~\ref{fig1}), which is affexted by a high reddening 
value. The similarity provides a consistent field subtraction, as suggested by the decontaminated 
resulting giant branches in Fig.~\ref{fig3}. Actually, differential absorption, when present, is more 
critical for low-density star clusters (e.g. the open cluster NGC\,6611, \citealt{N6611}), especially 
in the optical.

\subsection{Colour-magnitude filters}
\label{CMF}

Decontaminated CMDs present better defined cluster sequences. Based on these sequences, more intrinsic
colour-magnitude filters can be designed to remove stars with colours compatible with those of the
foreground/background field which, in turn, improves the cluster/background contrast. They are wide enough
to accommodate the colour distribution
of cluster CMD sequences, allowing for $1\sigma$ uncertainties. However, residual field stars with colours
similar to those of the cluster are expected to remain within the colour-magnitude filter. This residual 
contamination is statistically evaluated by means of the comparison field. Colour-magnitude filter widths 
should also account for formation or dynamical evolution-related effects, such as enhanced fractions of 
binaries (and other multiple systems) towards the central parts of clusters, since such systems tend to 
widen the sequences (e.g. \citealt{BB07a}; \citealt{N188}; \citealt{HT98}; \citealt{Kerber02}).

The colour-magnitude filters for the present GCs are shown in Figs.~\ref{fig1} - \ref{fig3} as a 
shaded region superimposed on the field-star decontaminated CMDs. In the present cases, they 
basically isolate giant-branch stars. As a rule, the filters are wide enough to take into account 
all the relevant stars in both colours. Thus, we only apply the method to the \jh\ colour.

\begin{figure}
\resizebox{\hsize}{!}{\includegraphics{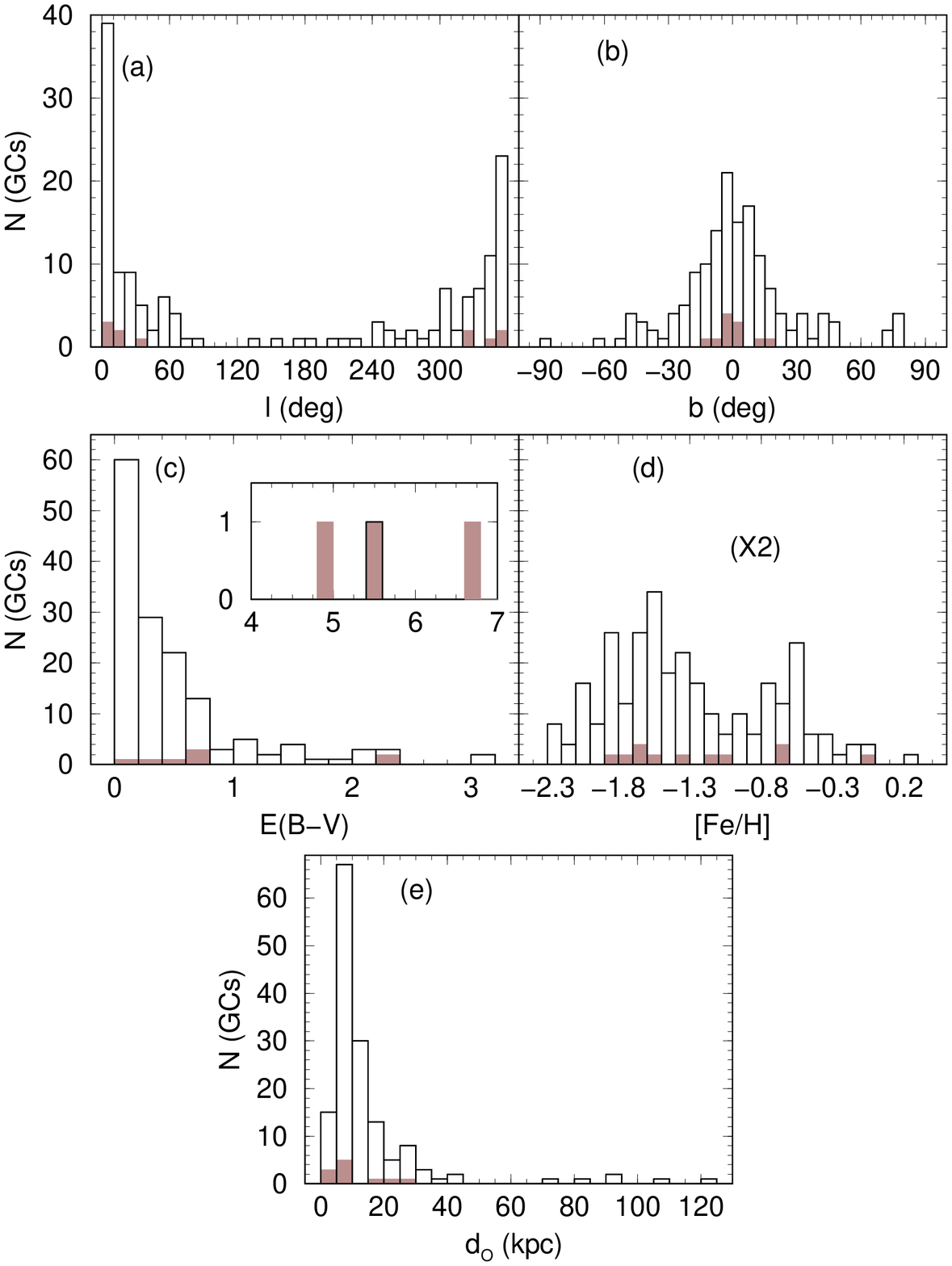}}
\caption{Astrophysical parameters of the present faint GCs (shaded histograms) compared to
the GCs in H03 (empty histograms). Inset of panel (c): GCs with high reddening values. The
y-scale in panel (d) was multiplied by 2 for clarity purposes.}
\label{fig7}
\end{figure}

\subsection{Absolute magnitude estimates}
\label{AbsMag}

ESO\,280-SC06, 2MASS-GC01 and 2MASS-GC02 lack absolute magnitudes, while the available estimate for
GLIMPSE-C01 has a large uncertainty (Table~\ref{tab1}). Here, such values are estimated following a 
similar method as that used for the recently identified GC FSR\,1767 (\citealt{FSR1767}). Basically, 
we compare the number of giant stars in ESO\,280-SC06, 2MASS-GC01, 2MASS-GC02 and GLIMPSE-C01 with 
that in the nearby ($\ds\approx2.2$\,kpc), intermediate-metallicity ($\feh=-1.2$) and total absolute magnitude
$\mv=-7.2$ GC M\,4 (H03, and references therein). In all cases, including M\,4, the number of member 
giant stars is computed from decontaminated 2MASS photometry (Sect.~\ref{FSD}), considering 
stars in the region from the cluster center to the tidal radius.

We assume that the decontaminated giants in ESO\,280-SC06, 2MASS-GC01, 2MASS-GC02 and
GLIMPSE-C01 are roughly of the same spectral type as in M\,4, and also that the luminosity of a
typical GC is dominated by the number of  giants. Thus, the total absolute magnitude of a given
GC can be estimated by scaling the total number of giants with that in M\,4. We found $\mv=-5.8\pm0.2$
for 2MASS-GC01, $\mv=-5.6\pm0.2$ for 2MASS-GC02, $\mv=-4.9\pm0.3$ for ESO\,280-SC06, and $\mv=-6.1\pm0.2$
for GLIMPSE-C01. Our estimate puts GLIMPSE-C01 $\approx2$\, mag fainter than that of \citet{KMB2005},
but still within their uncertainty. As a caveat we remark that the uncertainties in \mv\ only reflect
the statistical Poisson fluctuation of the number of giants in the above GCs and M\,4. They do not
take into account the individual spectral types or the presence of horizontal-branch stars (that are
clearly seen in the 2MASS CMD of the central region of M\,4 - \citealt{FSR1767}). Thus, they should be
taken as lower-limits to the actual uncertainties, especially in the case of ESO\,280-SC06, that is
$\approx22$\,kpc distant from the Sun. 2MASS-GC01, 2MASS-GC02 and GLIMPSE-C01 are only $\approx1$\,kpc
more distant from the Sun than M\,4.

The resulting RDPs for the sample GCs (Sect.~\ref{Struc}) follow the King-like profiles, which 
suggests that star-count losses in the central regions are not important. Similar analysis applied 
to M\,4 implies a loss of 2-3 giants only in the innermost region. Thus, the errors associated with
such effects are encompassed by the uncertainties quoted in Table~\ref{tab1} (col.~7).

In any case, the present total \mv\ estimates of ESO\,280-SC06, 2MASS-GC01, 2MASS-GC02 and GLIMPSE-C01
are consistent with the expected low-luminosity nature of these GCs (Sect.~\ref{Disc}).

\section{Structural parameters}
\label{Struc}

Structural parameters are derived using both the stellar radial number-density and surface brightness 
profiles, extracted from colour-magnitude filtered photometry (Sect.~\ref{CMF}).

To avoid oversampling near the centre and undersampling at large radii, the profiles are built in 
rings of increasing width with distance to the centre. The number and width of rings are adjusted 
to produce profiles with adequate spatial resolution and as small as possible $1\sigma$ Poisson 
errors. The residual background level of each RDP corresponds to the average number-density of 
colour-magnitude filtered stars measured in the comparison field. The $R$ coordinate and respective 
uncertainty in each ring correspond to the average position and standard deviation of the stars inside 
the ring. The vertical error bars correspond to the $1\sigma$ Poisson fluctuation of the star
counts in each ring. Obviously, the precision in the RDP level depends directly on the 2MASS resolution and 
photometric limit. Stars fainter than $\jj\approx17$ and/or those with projected companions close 
enough to produce distorted images have been discarded by 2MASS.

SBPs in the 2MASS \jj\ band are built similarly to the RDPs. The average surface brightness measured 
in the comparison field is subtracted from that computed in each cluster ring. As a caveat we note 
that the radial surface brightness is computed by summing the individual fluxes of all stars in each 
ring that have been resolved by 2MASS. Thus, SBPs are subject essentially to the same technical 
limitations as the RDPs. In this sense, the present SBPs are somewhat different (fainter) from the 
classical ones built with aperture photometry. 

By definition, RDPs represent star count densities, which do not take into account individual stellar 
magnitudes or asymmetries in the spatial distribution of stellar spectral types. Consequently, depending 
on the degree of large-scale mass segregation, RDP-derived parameters are expected to result larger than 
their counterparts obtained from SBPs.

\begin{figure}
\resizebox{\hsize}{!}{\includegraphics{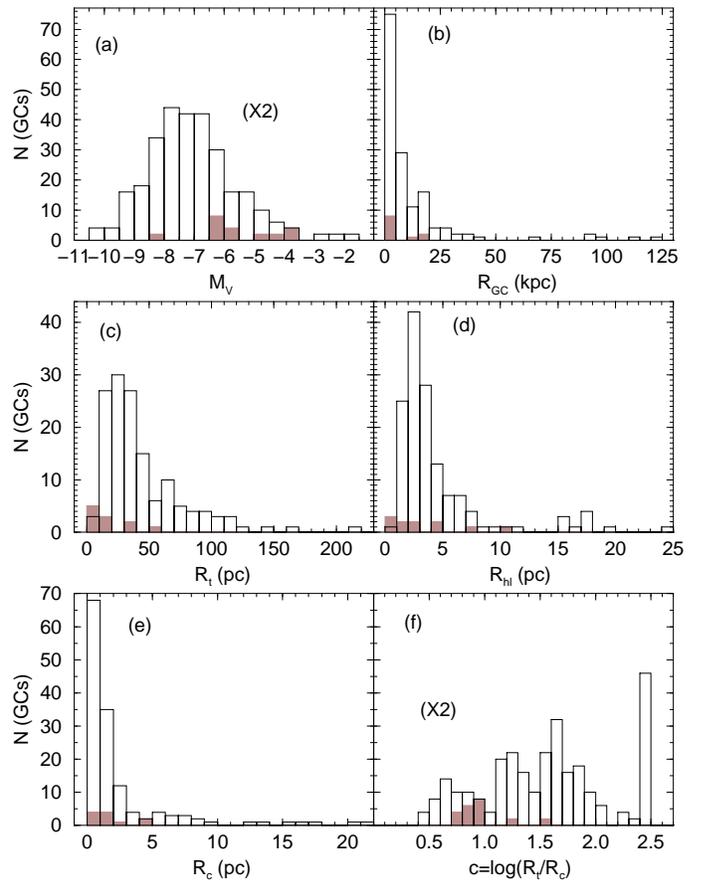}}
\caption{Same as Fig.~\ref{fig7}, but including the structural parameters (panels c - f). In panel (a) we
used $\mv=-8.4$ (\citealt{KMB2005}) for GLIMPSE-C01; our estimate is more compatible with a low-luminosity
GC, $\mv=-6.1\pm0.2$. Post-core collapse GCs in H03 occupy the $c=2.5$ bin in panel (f). The y-scale in panels
(a) and (f) was multiplied by 2 for clarity purposes.}
\label{fig8}
\end{figure}

The resulting RDPs are displayed in Fig.~\ref{fig4}, which also shows how the residual background 
contamination varies in each case. Because 2MASS is magnitude-limited, the contrast with the background
level and error bars among the RDPs are different and depend on the relative richness, distance from 
the Sun and direction in the Galaxy of each GC. Nevertheless, most of the RDPs are well-defined with
relatively high cluster/background contrasts, except for {bf AL\,3}, the most centrally projected GC of 
the present sample. 

We fit the SBPs with the empirical isothermal sphere, single-mass profile introduced by \cite{King1962}
to describe the surface brightness profiles of globular clusters. To the RDPs we applied the King-like
profile

\begin{equation}
\label{eq1}
\sigma(R)=\sigma_{bg}+\sigma_0\left[\frac{1}{\sqrt{1+(R/R_c)^2}} -
\frac{1}{\sqrt{1+(R_t/R_c)^2)}}\right]^2,
\end{equation}

where $\sigma_{bg}$ is the residual background density, $\sigma_0$ is the central number-density of
stars, and \rc\ and \rt\ are the core and tidal radii. This function is similar to \cite{King1962}
profile. In recent years, more sophisticated models have been employed to fit SBPs of Galactic and
extra-Galactic GCs. The most often used are the single-mass, modified isothermal sphere of \citet{King66}
that is the basis of the Galactic GC parameters given by \citet{TKD95} and H03, the modified isothermal
sphere of \citet{Wilson75}, that assumes a pre-defined stellar distribution function (which results in
more extended envelopes than \citealt{King66}), and the power-law with a core of \citet{EFF87} that has
been applied to massive young clusters especially in the Magellanic Clouds (e.g. Mackey \& Gilmore 2003a,b,c).
Each function is characterized by different parameters that are somehow related to cluster structure.
However, considering that the 2MASS photometric range covers basically the giant branch in most of the
present GCs, and that RDP and SBP error bars are significant, we decided for the simplest model
(\citealt{King1962}) as fit function.

In all cases the adopted King-like profile (Eq.~\ref{eq1}) represents well the RDPs along the full
radial scale (Fig.~\ref{fig4}), within uncertainties. However, the innermost point in the RDPs of Tz\,4
and 2MASS-GC01 presents a $1\sigma$ excess density over the fit, which might suggest post-core collapse.
GCs with that structure occur mostly in the bulge, but some halo GCs like NGC\,6397 also have it (e.g.
\citealt{CheDj89}; \citealt{TKD95}). Although to a lesser extent, the same occurs for ESO\,452-SC\,11.
Because the core radius results from the fit of Eq.~\ref{eq1} to the several radial points contained 
in RDPs, the probable post-core collapse structure of Tz\,4 and 2MASS-GC01 are not reflected on the
respective concentration parameters (Table~\ref{tab2} and Fig.~\ref{fig8}).

Two additional parameters are computed, the half-light radius (\rh), which can be used for comparisons 
with previous works, and the equivalent half-star count radius (\rhSC). The half-light radius is derived 
through direct integration of the background-subtracted SBPs, with no fit involved in the process. Derivation
of the distance from the center which contains half of the background-subtracted cluster stars, \rhSC, follows
a similar procedure, where number-densities are used instead of surface-brightness.

Fig.~\ref{fig5} displays the present RDPs in the usual, background-subtracted presentation
for GCs, which also shows the respective King-like fits and $1\sigma$-fit uncertainties.

RDP and SBP structural parameters that resulted from the above fits are given in Table~\ref{tab2}, 
in angular units, where we also provide the arcmin to parsec scale for each GC, based on the distances 
from the Sun listed in Table~\ref{tab1}. 

Within uncertainties, both types of profiles produce similar parameters. This is confirmed in
Fig.~\ref{fig6}, where RDP and SBP parameters (in absolute units) are compared. The agreement
is good for large radii, but fails for some GCs with small radii, especially the tidal one, where
RDP radii are larger than SBP ones. The best overall agreement occurs for the half-type radii, and
the approximate identity holds for about one order of magnitude in the radii scales (Sect.~\ref{Disc}).

\begin{table*}
\caption[]{Structural parameters derived from 2MASS data}
\label{tab2}
\renewcommand{\tabcolsep}{2.35mm}
\renewcommand{\arraystretch}{1.25}
\begin{tabular}{cccccccccccc}
\hline\hline
&&&\multicolumn{4}{c}{RDP}&&\multicolumn{4}{c}{SBP}\\
\cline{4-7}\cline{9-12}\\
GC&$1\arcmin$&&\rc&\rhSC&\rt&c&&\rc&\rhL&\rt&c\\
&(pc)&&(\arcmin)&(\arcmin)&(\arcmin)&&&(\arcmin)&(\arcmin)&(\arcmin)\\
(1)&(2)&&(3)&(4)&(5)&(6)&&(7)&(8)&(9)&(10)\\
\hline
IC\,1257     &7.3&&$0.2\pm0.1$&$1.4\pm0.3$&$7.1\pm2.0$&$1.5\pm0.2$&&$0.3\pm0.1$&$1.4\pm0.2$&$10.7\pm2.8$&$1.6\pm0.2$\\
Lyng\"a\,7   &2.1&&$0.9\pm0.2$&$1.2\pm0.1$&$8.4\pm1.2$&$1.0\pm0.1$&&$0.9\pm0.3$&$1.2\pm0.3$&$7.7\pm1.4$&$0.9\pm0.2$\\
Tz\,4        &2.6&&$0.8\pm0.4$&$1.9\pm0.2$&$6.2\pm1.4$&$0.9\pm0.2$&&$1.0\pm0.4$&$1.8\pm0.2$&$7.6\pm1.3$&$0.9\pm0.2$\\
Tz\,10       &1.7&&$1.0\pm0.3$&$1.6\pm0.2$&$5.9\pm2.0$&$0.8\pm0.2$&&$0.8\pm0.4$&$1.5\pm0.2$&$3.4\pm0.8$&$0.7\pm0.3$\\
BH\,176      &4.5&&$0.9\pm0.2$&$1.0\pm0.1$&$7.6\pm1.1$&$0.9\pm0.1$&&$0.9\pm0.2$&$0.8\pm0.1$&$5.9\pm0.5$&$0.8\pm0.1$\\
ESO\,452-SC11&2.3&&$0.5\pm0.2$&$0.5\pm0.2$&$5.4\pm1.8$&$1.0\pm0.2$&&$0.5\pm0.2$&$0.5\pm0.2$&$4.8\pm1.2$&$1.0\pm0.2$\\
ESO\,280-SC06&6.3&&$0.7\pm0.2$&$1.1\pm0.2$&$5.3\pm0.9$&$0.9\pm0.1$&&$0.5\pm0.2$&$1.0\pm0.2$&$5.0\pm1.1$&$0.9\pm0.2$\\
2MASS-GC01   &0.9&&$1.0\pm0.2$&$1.7\pm0.2$&$7.5\pm1.3$&$0.9\pm0.1$&&$0.7\pm0.2$&$1.6\pm0.2$&$4.4\pm0.4$&$0.8\pm0.2$\\
2MASS-GC02   &1.1&&$0.6\pm0.1$&$0.6\pm0.2$&$4.8\pm0.6$&$0.9\pm0.1$&&$0.4\pm0.1$&$0.5\pm0.2$&$3.7\pm0.4$&$1.0\pm0.1$\\
GLIMPSE-C01  &1.1&&$0.4\pm0.1$&$0.8\pm0.1$&$8.0\pm1.0$&$1.3\pm0.1$&&$0.4\pm0.1$&$0.5\pm0.1$&$4.5\pm0.5$&$1.1\pm0.1$\\
AL\,3        &1.7&&$0.4\pm0.3$&$0.5\pm0.2$&$4.1\pm0.5$&$1.0\pm0.3$&&$(\dag)$&$0.6\pm0.2$&$(\dag)$&$(\dag)$\\
\hline
\end{tabular}
\begin{list}{Table Notes.}
\item Col.~2: arcmin to parsec scale. Cols.~4 and 8: half-star count and half-light radii, 
respectively. Cols.~6 and 10: concentration parameter, $c=\log(\rt/\rc)$. $(\dag)$: The SBP 
of AL\,3 could not be fitted because of exceeding noise, especially beyond $R=1\arcmin$.
\end{list}
\end{table*}

Some structural parameters, derived from SBPs, are available for GLIMPSE-C01 (\citealt{KMB2005};
\citealt{IKB05}). The present values of core and half-light radii (Table~\ref{tab2}) are comparable
with those given in \citet{KMB2005}, and a factor $\sim2$ smaller than the core radius estimated 
by \citet{IKB05}. The present tidal radius, on the other hand, is $\sim1/6$ of that given by
\citet{IKB05}, which they estimated by extrapolation.

\section{Discussion}
\label{Disc}

\subsection{Photometric depth effects}
\label{PhotDepthEff}

The structural parameters dealt with in this paper were derived from RDPs and SBPs built with 
depth-limited photometry that, for the present GCs, correspond essentially to the giant branch. 
Potential effects on cluster radii introduced by this observational limitation should be taken
into account before comparing the parameters of present 11 faint GCs with those given in the
literature for the remaining GCs that, in general, have been obtained with deeper photometry.

This issue has been investigated by \citet{BB07b} by means of model star clusters of different
ages, structure and mass functions, with a spatial distribution of stars assumed to follow a 
pre-established analytical function. Model SBPs were built for the 2MASS \jj\ band. Cluster radii 
were then measured in RDPs and SBPs extracted from data with different photometric depths,
varying from the giant branch to the low-main sequence. The main results drawn from the GC
models are:

\begin{itemize}
\item {\em (i)} Structural parameters derived from SBPs are essentially insensitive to 
photometric depth.

\item {\em (ii)} RDPs built with shallow photometry yield cluster radii underestimated with 
respect to the values obtained with deep photometry. Tidal, half-star count and core radii 
are affected with increasing intensity for a given photometric depth. Tidal, half-star count 
and core radii underestimation factors are $\approx10\%$, $\approx15\%$ and $\approx25\%$,
respectively, for RDPs built essentially with giant-branch stars.

\item {\em (iii)} Profiles with photometry deeper than the turnoff have RDP radii systematically 
larger than SBP ones, especially the core. In the deepest profiles, RDP to SBP radii ratios are 
$\approx1.2$ both for the tidal and half-type radii, and $\approx1.4$ for the core radius.

\item {\em (iv)} Changes in RDP parameters with photometric depth result from a spatially variable
mass function. 
\end{itemize}

\begin{figure}
\resizebox{\hsize}{!}{\includegraphics{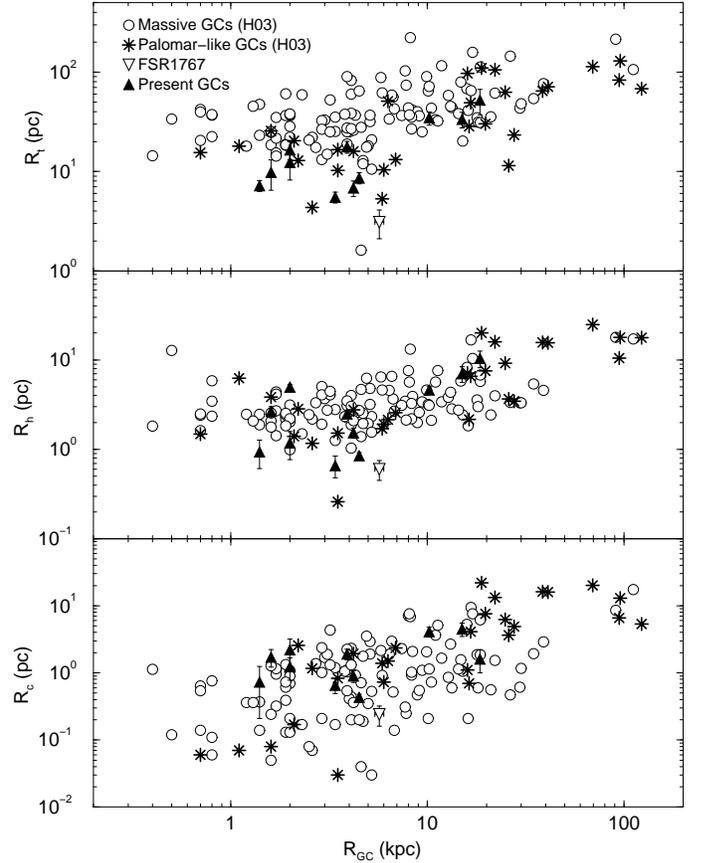}}
\caption{Dependence of cluster radii on Galactocentric distance. Empty circles: massive ($\mv<-6$) 
GCs in H03. Asterisks: Palomar-like ($\mv\protect\ga-6$) GCs from H03. Empty triangle: FSR\,1767.
Filled triangles: present faint GCs.}
\label{fig9}
\end{figure}

Most of the above results are directly related to cluster dynamical evolution. Large-scale mass segregation 
drives preferentially low-mass stars towards large radii (while evaporation pushes part of these 
stars beyond the tidal radius, into the field), and high-mass stars towards the central parts of 
clusters. If the stellar mass distribution can be described by a spatially variable mass function 
flatter at the center than in the halo, the resulting RDP radii should be larger than SBP ones. 
The differences should be more noticeable in the core than the half and tidal radii, since the 
core would contain, on average, stars more massive than the halo, and especially near the tidal 
radius. Such differences, in addition, increase for profiles built with deeper photometry
(\citealt{BB07b}).

It follows from the above that radial number-density and surface-brightness profiles built with
shallow photometry, as is the case of the present GCs, are expected to yield similar values for
the three types of cluster radii, and that RDP radii are underestimated by the factors given in
{\em (ii)} above with respect to the intrinsic values. However, we note that the underestimation factors
for the present faint GCs are, in most cases, smaller than (or similar to) the $1\sigma$ fit
uncertainties (Table~\ref{tab2}). The similarity between the values of RDP and SBP radii effectively 
occurs for most of the present faint-GC sample (Fig.~\ref{fig6}), especially concerning the half-light
and half-star count radii.

\subsection{Comparison with GCs in H03}
\label{CompHo3}

In Fig.~\ref{fig7} we compare the present faint GCs with the GCs in H03 in terms of Galactic coordinates
(panels a and b), reddening \ebv\ (panel c), metallicity \feh\ (panel d), and distance from the Sun
(panel e). The present GCs are projected not far from the bulge and close to the plane. Such
difficult directions, together with relatively high reddening values in the optical (GLIMPSE-C01,
2MASS-GC01 and 2MASS-GC02 have $A_V\ga15$), explain why the present GCs have so far been scarcely
studied. They have metallicities as low as $\feh=-1.8$ with a single case (BH\,176) being metal-rich
($\feh=0.0$), and $\approx73\%$ have metallicity lower than $\feh=-1.0$. The same fraction of GCs is 
less than $\approx10$\,kpc distant from the Sun.

Further comparisons with the GCs in H03 are explored in Fig.~\ref{fig8}. With respect to luminosity, 
most of the GCs of the present sample are $\approx3$\,mag fainter than the 
average luminosity of the H03 GC distribution, while at least 3 of them populate the faint-magnitude 
tail (panel a). The exception perhaps is GLIMPSE-C01 with $\mv=-8.4$, but with $\approx3$\,mag of 
uncertainty (\citealt{KMB2005}). Our estimate gives $\mv=-6.1\pm0.2$ for GLIMPSE-C01, which is 
consistent with a low-luminosity GC. About 73\% of them are located within 5\,kpc of the Galactic 
center (panel b), which implies that some are probably bulge GCs; the remaining GCs (IC\,1257,
BH\,176 and ESO\,280-SC06) have Galactocentric distances compatible with the halo. The present 
tidal radius distribution suggests a bias towards small values (panel c) with respect to H03, 
while the half-star count radii are more evenly distributed (panel d). The core radii appear to 
be distributed similarly to those in H03 (panel e). Consequently, the concentration parameters 
(panel f) are biased towards GCs of loose structure. Loose structure together with low-luminosity 
($\mv\ga-6$) are typical of Palomar-like GCs, i.e. low-mass GCs that do not contain populous giant
branches (\citealt{FSR1767}, and references therein).

\subsection{Dependence on Galactocentric distance}
\label{DGD}

Finally, in Fig.~\ref{fig9} we examine the dependence of cluster radii on Galactocentric distance.
GCs from H03 shown in this figure are separated according to luminosity, massive ($\mv<-6$) and
Palomar-like ($\mv\ga-6$) GCs. We also include for comparison the recently identified Palomar-like 
GC FSR\,1767 (\citealt{FSR1767}). Massive and Palomar-like GCs follow the well-known relation of 
increasing cluster radii with Galactocentric distance (e.g. \citealt{MvdB05}; \citealt{DjMey94}).
However, compared to the massive GCs, the Palomar-like ones tend to have smaller tidal and larger 
core radii than the massive ones for a given Galactocentric distance, especially for small \Rgc,
which is consistent with their loose structures. The half-light radii distributions of both
types of GCs are very similar.

Concerning the 11 faint GCs dealt with in this paper, the core and tidal radii distributions
with Galactocentric distance are similar to those of the Palomar-like GCs. We remark that this
conclusion would suffer only minor changes in the case of correcting the measured radii for the 
photometric depth effects discussed above (Sect.~\ref{PhotDepthEff}). The intrinsic (i.e. corrected)
core radii would be somewhat larger than the measured ones. To a lesser extent, the same applies
to the relation between depth-limited and intrinsic tidal radii, thus confirming the loose structure
of the present GCs.

Associated with the low-luminosities (Table~\ref{tab1}), the radii distributions with Galactocentric
distance imply that the present faint GCs share similar characteristics as the sample of Palomar-like
GCs present in H03.

\section{Concluding remarks}
\label{Conclu}

Compilations of astrophysical parameters of globular (e.g. H03) and open (e.g. WEBDA\footnote{\em
obswww.unige.ch/webda}) clusters represent a fundamental step in the study of several aspects
related to the clusters themselves and the Galactic subsystems they belong to.

The majority of the GCs are rather populous and concentrated, which results in high contrast with
respect to the background/foreground stellar field. The bottom line is that such GCs can be rather
easily observed in most regions of the Galaxy, from near the center to the remote halo outskirts.
Thus, through the statistics of observables such as the age, metallicity, luminosity, dynamics and
kinematics, GCs can be taken as excellent tracers of the formation history of the Galaxy. Obviously,
the same applies to extra-Galactic GCs and their host galaxies.

Faint globular clusters, in particular, are important because, compared to massive clusters, they
are more severely affected by the many Galaxy tidal effects and have shorter dynamical-evolution
time scales. Consequently, the distribution of their structural parameters throughout the Galaxy
may provide clues to better understand the dynamical evolution of individual GCs and their system.

With the present work we derived structural parameters of 9 faint GCs listed in H03, and 2 others
recently identified, by means of stellar radial number-density and surface-brightness profiles.
These GCs either lacked such parameters completely or, in some cases, had first-order estimates
based on CCD observations covering fields smaller than the angular diameters of clusters. In the
present work, the profiles were built with wide-field 2MASS photometry (\jj\ band), whose depth-limit
together with the distances of the present GCs, result basically in giant-branch stars. In all cases,
the radial profiles reached at least to the respective tidal radius. We also estimated total \mv\
values, which lacked for ESO\,280-SC06, 2MASS-GC01 and 2MASS-GC02, and provided an independent \mv\
estimate for GLIMPSE-C01 (which resulted $\approx2$\,mag fainter than previously given), based on
field-star decontaminated photometry of the giant stars, using M\,4 as a template GC. Our estimates
are in the range $-6.1\la\mv\la-4.9$, which confirms them as low-luminosity GCs.

The stellar radial profiles were fitted with King-like functions from which the core, half-star
count, half-light and tidal radii, as well as the concentration parameter, were derived. Most of
the present objects turned out to be so-far neglected Palomar-like (low-mass, loose structure)
GCs. The distributions of core and tidal radii, with respect to Galactocentric distance, are also
consistent with the Palomar-like nature.

Besides a better understanding on the individual structure of some faint globular clusters, what
also results from this work is an improved coverage of the GC parameter space.

\begin{acknowledgements}
We thank the anonymous referee for helpful suggestions.
We acknowledge partial support from the Brazilian agency CNPq.
\end{acknowledgements}


\end{document}